\documentclass[twocolumn,showpacs,preprintnumbers,superscriptaddress]{revtex4}
\usepackage{amsmath}
\usepackage{amssymb}
\usepackage{amsmath}
\usepackage{epsfig}
\usepackage{graphicx}
\usepackage{inputenc}
\inputencoding{latin1}
\usepackage{ulem}

\ifx\hypersetup\sadfkjashdfkxja\else
\hypersetup{pdfsubject={}}
\hypersetup{pdfkeywords={}}
\hypersetup{plainpages=false}
\hypersetup{bookmarksnumbered=true}
\hypersetup{pdfstartview=FitH}
\hypersetup{pdfpagemode=UseNone}
\hypersetup{colorlinks=false}
\hypersetup{citebordercolor={.5 1 .5}}
\hypersetup{urlbordercolor={.5 1 1}}
\hypersetup{linkbordercolor={1 .7 .7}}
\fi

\newcommand{\ud}{\mathrm{d}}
\newcommand{\be}{\begin{equation}}
\newcommand{\ee}{\end{equation}}
\newcommand{\bea}{\begin{eqnarray}}
\newcommand{\eea}{\end{eqnarray}}

\newcommand{\Appendix}[1]%
    {%
     \section{#1}%
      }



\begin{document}

\title{A method for getting a finite $\alpha$ in the IR region from an all-order beta function}
\author{Zhi-Yuan Zheng}
\email{zhengzy@itp.ac.cn}
\affiliation{Key Laboratory of Theoretical Physics, Institute of Theoretical Physics, Chinese Academy of Sciences\\ Beijing 100190, People's
Republic of China}
\affiliation{School of Physical Sciences, University of Chinese Academy of Sciences\\No.19A Yuquan Road, Beijing 100049, China}
\author{Gao-Liang Zhou}
\email{zhougl@itp.ac.cn}
\affiliation{College of Science, Xi'an University of Science and Technology, Xi'an 710054, People's Republic of China}
\begin{abstract}
The analytical method of QCD running coupling constant is extended to a model with an all-order beta function which is inspired by the
famous Novikov-Shifman-Vai\-n\-s\-htein-Zakharov beta function of N=1 supersymmetric gau\-g\-e theories. In the approach presented here, the running coupling is determined by a transcendental equation with non-elementary integral of the running scale $\mu$. In our approach $\alpha_{an}(0)$, which reads 0.30642, does not rely on any dimensional parameters. This is in accordance with results in the literature on the analytical method of QCD running coupling
constant. The new ``analytically im\-p\-roved'' running coupling constant is also compatible with the property of asymptotic freedom.
\end{abstract}
\maketitle
\section{Introduction.}
\label{introduction}
The renormalization scale dependence or evolution of running coupling constant is described by the beta
function of a theory concerned. The explicit expression for beta function can be obtained by means of perturbation methods, and the validity of its use depends on whether its resultant coupling constant is excessively
large or not. Solving a given beta function, we can obtain an expression for running coupling constant, which may blow up or become enough large somewhere, indicating that the perturbation breaks down and the invalidity of using this given beta function. This problem leads us to expect that, as has been done in Ref. \cite{Shirkov:1997wi}, running coupling constant must cease to increase somewhere (i.e. freeze somewhere).

The freezing of running coupling constant (FORCC) is closely related to the explicit form of beta function. Using beta function to study the
FORCC is meaningful and direct; the freezing of running coupling constant might
appear as a result of beta function's vanishing---FORCC is related to infrared or ultraviolet zero of beta function. In Ref. \cite{Banks:1981nn} infrared zero of beta function was studied at three-loop order. In Ref. \cite{Gardi:1998rf,Gardi:1998ch,Chishtie:1999tx,Ryttov:2010iz,Shrock:2013pya,Shrock:2013ca,Pica:2010xq} this was carried out to four-loop order. Recently, in Ref. \cite{Ryttov:2016ner} this has been carried out to five-loop order.

The property of freezing coupling constant $\alpha$ at
low energy scale in QCD has been studied for many years \cite{Badalian:1997de,Shirkov:1997wi} and been widely used in QCD
phenomenology \cite{Eichten:1974af,Eichten:1979ms,Richardson:1978bt,Godfrey:1985xj,Mattingly:1993ej,Dokshitzer:1995ev,Badalian:1997de,Badalian:2001by}. The phenomenological
evidence for running coupling constant of QCD to freeze in IR region is numerous and increasing. Various models have been developed to investigate FORCC, and the prediction value for $\alpha_s(0)$ of QCD ranges from $0.4$ to $1$ in phenomenology \cite{Godfrey:1985xj,Mattingly:1993ej,Mattingly:1992ud,Aguilar:2001zy,Dokshitzer:1995zt,Dokshitzer:1995qm}. Also, there are theoretical reasons in favor of
FORCC \cite{Banks:1981nn,Aguilar:2001zy}. Even though the behaviour of running coupling of QCD in IR region is nonperturbative in usually sense, in Ref. \cite{Mattingly:1993ej} it was shown that perturbation theory provided some theoretical evidence for FORCC.

Many methods have been used to investigate FORCC. For example, in Ref. \cite{Roberts:1994dr} Schwinger-Dyson equation was used to investigate FORCC. Another well known way to study FORCC is to study coupling constant directly---we can obtain the knowledge of running coupling in IR region from the knowledge of it in UV region \cite{Shirkov:1997wi}. In Ref. \cite{Shirkov:1997wi}, through the use of ``analytization procedure'', an ``analytically-im\-p\-roved'' expression for running coupling was obtained, which is free of ghost-pole and has a universal limit value. The
 ``analytization procedure'' elaborated in Ref. \cite{Redmond:1958juf,Redmond:1958pe,trio} and used in Ref. \cite{Shirkov:1997wi} includes three steps:\\
(I) Finding an explicit expression for $\alpha_s(\mu^2)$ in the
Euclidean region;\newline
(II)Performing analytical continuation into the Minkowski region. Extracting its imaginary part for defining the spectral
density by $\rho_{RG} (\sigma ,\alpha)= \text{Im}\alpha(-\sigma
-i\epsilon ,\alpha)$;\newline
(III) Using $\rho_{RG}$ to define an ``analytically-improved''
running coupling constant in the Euclidean region.

In some sense these three steps are perturbative, since $\alpha$ is obtained by perturbation methods usually. An all-orders beta function was first proposed in Ref. \cite{Ryttov:2007cx}. A proof of this is given in Ref. \cite{Pica:2010mt}. In our work, this beta function is used to extract
 $\alpha_{an}(\mu)$ by a similar way as the way used in Ref. \cite{Shirkov:1997wi}.

The remainder of this paper is organised as follows. In section 2, we start with an all-orders beta function and concentrate on eliminating the ghost pole. In section 3, after having removed singularity, we get an equation from which we define a new ``analytically-im\-p\-roved'' running coupling which is free of ghost pole and does respects asymptotic freedom. In section 4, the issue of scheme dependence is discussed.
\section{ELIMINATION OF SINGULARITY}
\label{ELIMINATION OF SINGULARITY}

As has been said before, an all-order beta function was first proposed in Ref. \cite{Ryttov:2007cx} and proved in Ref. \cite{Pica:2010mt}, which takes the form
\begin{equation}
\frac{\beta(\alpha)}{\alpha}=-\frac{\alpha}{2\pi}\frac{\alpha+\sum_{r=1}^p\alpha_rN_r\gamma_r}{1-\frac{\alpha}{2\pi}\alpha_g} .
\end{equation}
For pure Yang-Mills theory this can be simplified to
\begin{equation}
\frac{\beta_{YM}(\alpha)}{\alpha}=-\frac{11}{3}\frac{\alpha}{2\pi}\frac{C_2[G]}{1-\frac{\alpha}{2\pi}\frac{17}{11}C_2[G]} .
\end{equation}
In SU(N) gauge theory, $C_2(G)$ is just N. Carrying out the obvious transformation $\frac{\alpha}{2\pi}\rightarrow\alpha$ converts Eq. (2) to a new compact beta function in the form
\begin{equation}
\beta_{YM}(\alpha)=-\frac{C_1\alpha^2}{1-C_2\alpha},
\end{equation}
where $C_1=11C_2(G)/3$ and $C_2=17C_2(G)/11$. The general solution of this Equation is of the form
\begin{equation}
C_2\ln\frac{\alpha(\mu)}{\alpha(\Lambda)}+(\frac{1}{\alpha(\mu)}-\frac{1}{\alpha(\Lambda)})=C_1\ln\frac{\mu}{\Lambda},
\end{equation}
where $\Lambda$ is just a integral constant. For later convenience, inverting Eq. (4) and making a replacement $\ln\frac{\mu}{\Lambda}\rightarrow\frac{1}{2}\ln\frac{\mu_E^2}{\Lambda_E^2}$ (here $\mu_E^2=-\mu^2$ and $\Lambda_E^2=-\Lambda^2$), we have
\begin{equation}
\frac{1}{C_2\ln\frac{\alpha(\mu)}{\alpha(\Lambda)}+(\frac{1}{\alpha(\mu)}-\frac{1}{\alpha(\Lambda)})}=\frac{2}{C_1\ln\frac{\mu_E^2}{\Lambda_E^2}}.
\end{equation}
For simplicity, we denote by $y_0(\mu_E^2)$ the right-hand side of Eq. (5) and by $y_0(\alpha(\mu))$ the left-hand side of Eq. (5).

The analyticity property of $y_0(\alpha(\mu))$ is almost the same as that of running coupling $\alpha(\mu)$ with the exception that the coupling
constant is non-positive or may be a pole of $y_0(\alpha(\mu))$. Therefore, in this work we shall concentrate on studying the analyticity properties of $y_0(\alpha(\mu))$ and $y_0(\mu_E^2)$. For simplicity, we set $\alpha(\Lambda)=1$
and $\lambda=C_2$, then
\begin{eqnarray}
y_0(\alpha(\mu))=\frac{1}{\frac{1}{\alpha(\mu)}+(\lambda\ln\alpha(\mu)-1)}.
\end{eqnarray}


It may be shown that the denominator of $y_0(\alpha(\mu))$ can be zero only if the running coupling $\alpha$ is a real number.

For the continuation of this work, we make some reasonable assumptions here; self-consistency of some of them can be verified by later check.

Let us make the first assumption: for the running coupling to be real-valued, what is necessary is that $\mu^2$ must be real-valued. Self-co\-n\-sistence of this assumption can be checked later.

The second assumption is the absence of Landau-pole. Therefore, according to asymptotic freedom, which will be discussed later, there is no point on the real axis at which $\alpha_s$ is negative. When $\mu_E^2$ goes to$-\infty$, according to asymptotic freedom, $\alpha(\mu)$ tends to $0$, which leads to the denominator of $y_0(\alpha(\mu))$ having a non-zero value. Therefore, the distance between points at which the denominator of $y_0(\alpha(\mu))$ vanishes and the origin must reach its maximum (if exist) at a point on the negative real axis---we set it to be $-\mu_0^2$.

As is well known, if a function is analytical in a connected region surrounded by a contour, we can use Cauchy theorem
\begin{equation}
\mathit{y}(\mu^2)=\frac{-i}{2\pi}\oint\frac{y(x)}{x-\mu^2}\ud x ,
\end{equation}
which in this paper is used to subtract singularity terms.

The third assumption is that the integral along infinite contour is summed to zero. Note that $y_0(\alpha(\mu))$ is analytical except
$\mu_E^2$ is a pole of it on the real axis. Therefore, in this paper, the integral is taken over the contour shown in Fig. 1. Thus we immediately arrive at
\begin{equation}
y_0(\mu_E^2)=\frac{1}{\pi}\int_{-\mu_0^2}^\infty\frac{\text{Im}y_0(x+i\epsilon)}{x-\mu_E^2}\ud x
\end{equation}
We note that
\begin{equation}
\text{Im}\ln\frac{\mathit{a}+i\epsilon}{\mathit{b}}=\begin{cases}
0,&\text{if}\,\mathit{a}<0\,\text{and}\,\mathit{b}<0;\\
-\pi,&\text{if}\,\mathit{a}>0\,\text{and}\,\mathit{b}<0 .
\end{cases}
\end{equation}
Thus Eq. (8), can be further simplified into
\begin{equation}
\mathit{y_0}(\mu_E^2)=\frac{2}{C_1}\int_0^\infty\frac{1}{x-\mu_E^2}\frac{\ud x}{\ln^2|\frac{x}{\Lambda_E^2}|+\pi^2}
\end{equation}
\begin{figure}
\includegraphics[scale=0.6]{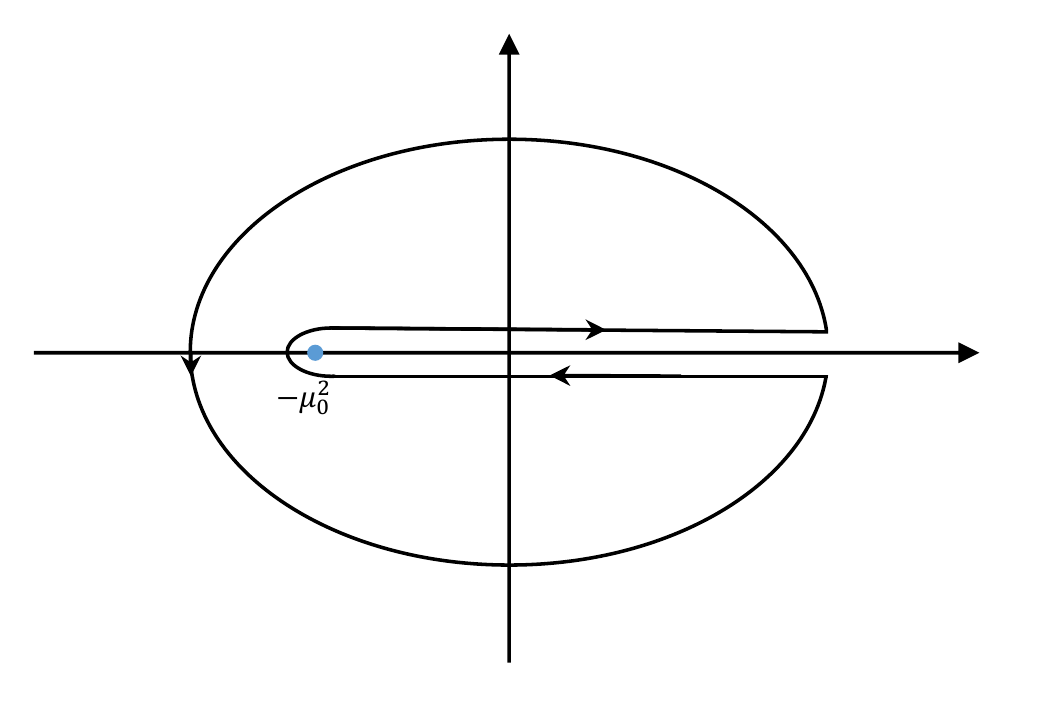}
\caption{Cauchy theorem:The horizontal axis represents the real part of $\mu_E^2$, and the vertical axis represents the imaginary part of $\mu_E^2$}
\end{figure}
\section{EVALUATION OF COUPLING CONSTANT}
\label{EVALUATION OF COUPLING CONSTANT}
It's almost always the case that the values of $y_0(\mu_E^2)$ are related to $\Lambda$. Two cases $\Lambda=200MeV$ and $\Lambda=400MeV$ are shown in Fig. 2. However, there is one obvious exception that $y_{an}(0)$ is obviously independent of the value of $\Lambda$. Therefore, the value for $\alpha_{an}(0)$ is independent of $\Lambda$.
\begin{figure}
\begin{center}
\includegraphics[width=0.41\textwidth]{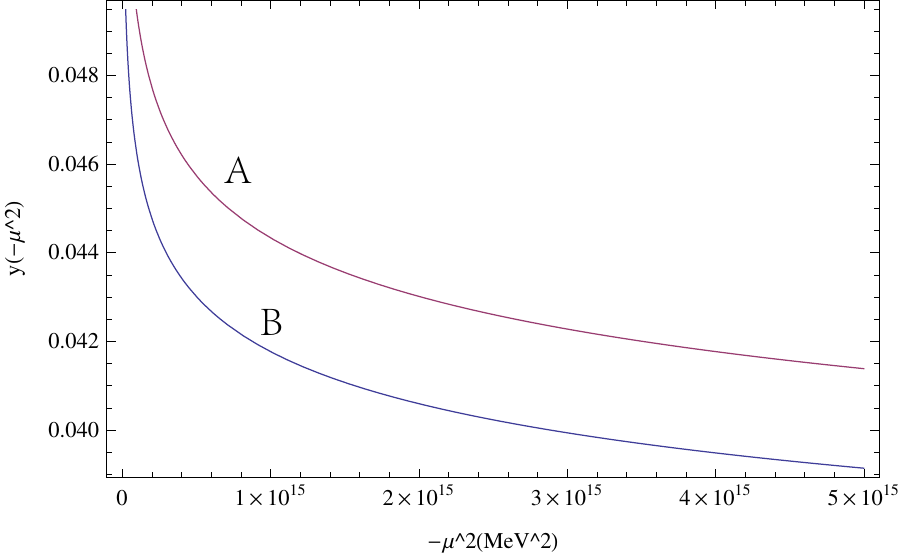}
\end{center}
\caption{The curve A and B are the ${y_{an}}(\mu^2)$ for $\Lambda=400MeV$ and $\Lambda=200MeV$ respectively}
\end{figure}

Since the first part of the integrand in Eq. (10) is a monotone decreasing function of energy scale $\mu$, it's obvious that, as can be seen form Fig. 2, $y_0(\mu_E^2)$ is a monotone decreasing function of energy scale.

By straightforward calculation, we obtain an explicit expression for Eq. (10) in the form
\begin{equation}
y_{an}(\mu)=\frac{2}{C_1}(\frac{1}{\ln\frac{\mu^2}{\Lambda^2}}+\frac{\Lambda^2}{\Lambda^2-\mu^2})
\end{equation}
which decreases monotonically from $2/C_1$ to zero as $\mu$ varies from zero to infinity, and is free of ghost-pole compared with the original expression $2/(C_1\ln\mu^2/\Lambda^2)$.

Acquisition of $\alpha_{an}(\mu^2)$, in our approach, is determined by equation
\begin{align}
y_{an}(\alpha_{an}(\mu))&=\frac{\alpha_{an}(\mu)}{[1+\alpha_{an}(\mu)(\lambda\ln\alpha_{an}(\mu)-1)]}\nonumber\\ &=\frac{2}{C_1}(\frac{1}{\ln\frac{\mu^2}{\Lambda^2}}+\frac{\Lambda^2}{\Lambda^2-\mu^2})
\end{align}
which is an ``analytically-im\-p\-roved'' equation corresponding to $y_0(\alpha(\mu))$=$y_0(\mu_E^2)$.

As can be seen from Fig. 3 the new ``analytically-im\-p\-roved'' running coupling $\alpha_{an}$ as a function of $\mu$ is two-valued: each value of the $y_{an}(\mu)$ corresponds to two different values of $\alpha_{an}$. The difference between this two choices of value for running coupling, as can be seen from Fig. 3, is obviously large and varies monotonously with energy scale. Therefore, the determination of running coupling is a little more complicated.

The property of $y_{an}(\alpha_{an}(\mu))$ as function of $\alpha_{an}(\mu)$ is somewhat complicated. Differentiating $y_{an}(\alpha_{an})$ with respect
to $\alpha_{an}$, we have
\begin{equation}
y'_{an}(\alpha_{an}(\mu))=\frac{1-\lambda\alpha_{an}(\mu)}{(1+\alpha_{an}(\mu)(\lambda\ln\alpha_{an}(\mu)-1))^2}
\end{equation}
whose numerator is $1-\lambda\alpha_{an}(\mu)$ and whose denominator is always nonnegative. We note that, as long as
$\alpha_{an}(\mu)$ does not exceed $\frac{1}{\lambda}$, $y'_{an}(\alpha_{an}(\mu))$ maintain its positivity. The appearances of abrupt change shown in Fig. 3 result from the vanishing of denominator of $y'_{an}(\alpha_{an}(\mu))$.
\begin{figure}
\begin{center}
\includegraphics[width=0.4\textwidth]{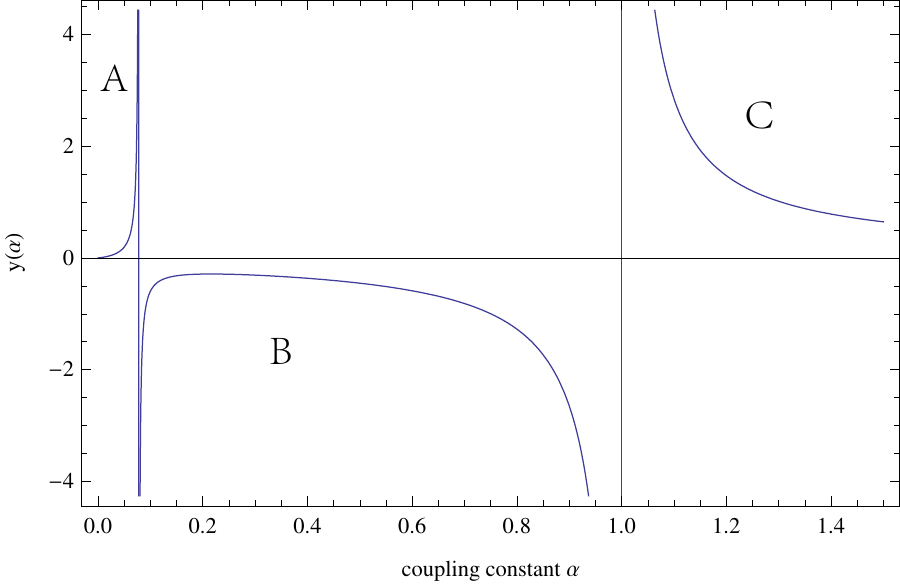}
\end{center}
\caption{Horizontal Axis represents coupling constant $\alpha_{an}$, Vertical Axis represents $y_{an}(\alpha_{an})$. Region B has no corresponding $y_{an}(\mu)$, and should not be taken into consideration later}
\end{figure}

It may be shown that, of the three regions represented by A, B and C respectively in Fig. 3, the middle one (represented by B) corresponds to un-physical region and therefore can be discarded, since $y_{an}(\mu)$, with an explicit expression shown in Eq. (11), is always positive. Since $y_{an}(\mu)$ is a monotone decreasing function of energy scale, of the remainder region A and C, the left region A in Fig. 3 corresponds to region obeying asymptotic freedom (as energy scale increases, $y_{an}(\mu)$ decreases which leads to the decrement of $\alpha_{an}(\mu)$), and the right region C in Fig. 3 corresponds to region violating asymptotic freedom (as energy scale increases coupling constant increases, what is worse, in UV region the coupling constant blows up). Taking our assumption of asymptotic freedom, physical intuition and mathematical continuity into consideration region C also can be discarded. The distinction between regions A and C also may be traced back to the explicit form of all-orders beta function in Eq. (3). As energy scale changes from IR region to UV region, evolution of coupling constant is essentially determined by the initial value of running coupling in IR region. If this value is smaller than $\frac{1}{C_2}$, asymptotic freedom is obeyed, otherwise is violated dramatically.

Now for a fixed $\mu$, taking all the discussion given above into consideration, we can fix the corresponding $\alpha_{an}(\mu)$. Even though, it's still impossible to acquire a expression for $\alpha_{an}(\mu)$ in terms of $\mu$ explicitly, numerical value for $\alpha_{an}$ can be obtained for any $\mu$. Letting $\mu$ vanish, for SU(3) gauge theory, we get a numerical value $0.0487688$ in the region A of Fig. 3 (region obeying asymptotic freedom) and a numerical value $3.8413$ in the region C of Figure.3 (region violating asymptotic freedom dramatically). Therefore, we choose 0.0487688 as our result. Since a transformation $\frac{\alpha}{2\pi}$ $\rightarrow$ $\alpha$ has been made early, the running coupling constant in IR region freezes at 0.30642.
\section{DISCUSSION OF SCHEME DEPENDENCE}
\label{DISCUSSION OF SCHEME DEPENDENCE}
As is well known, for a given theory to be renormalizable, we impose the requirement that the divergent part of any one-particle irreducible diagram can be cancelled by a local counterterm generated from the bare Lagrangian of this given theory. Therefore, once the divergences of a given theory have been subtracted out by a certain renormalization scheme, one is still free to perform further finite renormalization, thus resulting in the may-be scheme dependence of quantities in field theory.  Here in this section, our discussion is restricted within the class of mass-independent renormalization scheme.

As far as we are concerned here, for beta function, usually only the first two none-vanishing coefficients of the expansion for beta function are scheme independent. There is one obvious may-be exception to this statement just given above. If, for a given theory, an all-orders beta function can be obtained in a certain renormalization scheme and written in the form
\begin{equation}
\beta(\alpha)=\frac{\beta_1\alpha^n}{1-\frac{\beta_2}{\beta_1}\alpha^m},
\end{equation}
and if this form can be maintained in anther scheme, it's obvious that the all-orders beta functions in this two schemes are equal to each other.

The expression for beta function expanded in terms of running coupling is usually of the form
\begin{equation}
\beta(\alpha)=\frac{\ud \alpha}{\ud \ln\mu}=\beta_0\alpha^{n_0}+\beta_1\alpha^{n_1}+\beta_2\alpha^{n_2}+\ldots,
\end{equation}
which can be rewritten as
\begin{equation}
\beta(\alpha)=\frac{\ud \alpha}{\ud \ln\mu}=\frac{\beta_0\alpha^{n_0}}{1+\beta'_1\alpha^{n'_1}+\beta'_2\alpha^{n'_2}+\ldots}.
\end{equation}
These new coefficients of Eq. (16) can be obtained by matching with Eq. (15) order by order.

Along the same procedures carried out above, we may obtain an equation between the new ``analytically-im\-p\-roved'' coupling $\alpha_{an}$ and energy scales in the form
\begin{equation}
\frac{1}{a_0\ln \alpha_{an}(\mu)+\sum_{i\neq0} a_i\alpha_{an}(\mu)^i+C}=2(\frac{1}{\ln\frac{\mu^2}{\Lambda^2}}+\frac{\Lambda^2}{\Lambda^2-\mu^2}),
\end{equation}
where $\Lambda$ is just an integral constant and C also a constant. It's obvious that, if $\alpha_{an}$ is in the perturbation region (i.e. small enough), of all the terms in the denominator of the expression in the left-hand side of Eq. (17), the term with a lower index $i$ contribute more (the first two none-vanishing term with scheme independent coefficients contribute most).

To conclude this section, we may make the following remarks concerning the scheme dependence of the new ``anal\-y\-t\-ically-im\-p\-roved'' freezing coupling. First, in our approach the existence of freezing coupling is independent of renormalization scheme we choose, though the explicit value is scheme dependent. Second, if the running coupling is in perturbation region, the freezing coupling is effectively (within the accuracy required) determined by a finite number of none-vanishing coefficients of beta function.
\section{CONCLUSION}
\label{CONCLUSION}
The appearance of ghost pole in the expression for running coupling of QCD or QED, leads to the hypothesis that the running coupling must freeze somewhere. In this article, starting with an all-orders beta function, we have obtained an `anal\-y\-t\-ically-im\-p\-roved'' coupling by some analyticity procedure to remove singularity. This new running coupling is free of ghost pole and does respect asymptotic freedom.

Many works have been done to investigate FORCC theoretically or phenomenologically. It is clear that much more work has to be done in
order to get a profound understanding of this amazing property which is essential to physics.
\section*{ACKNOWLEDGEMENT}
We thank Prof. Y. Q. Chen and Dr. P. Wu for helpful discussions and important suggestions on the manuscript. One of us (Z. Y. Zheng) thanks Y. Z. Xu and Z. L. Cui for giving necessary helps to finish this work. The work of Z. Y. Zheng is supported by The National Nature Science Foundation of
China under Grant No. 11275242. The work of G. L. Zhou is supported by The National Nature Science Foundation of China under Grant No. 11647022.
\appendix
\section{}
In this Appendix we shall prove a statement given above, and check an assumption made above.

Firstly, we prove the statement that the denominator of $y_0(\alpha(\mu))$ shown in Eq. (6) can be zero only if $\alpha(\mu)$ is real number.

Let's set $\alpha=re^{i\theta}$. Then the condition for the denominator of $y_0(\alpha(\mu))$ to vanish can be expressed as
\begin{align}
\lambda\theta &=\frac{1}{r}\sin(\theta) ,\\
1&=\lambda\ln r+\frac{\cos(\theta)}{r} .
\end{align}
From condition (A.1), we have
\begin{equation}
\lambda r\leq1 ,
\end{equation}
which leads to (Note that for SU(3), $\lambda=51/11$)
\begin{equation}
\cos(\theta)>0 .
\end{equation}
Now we can set $-\frac{\pi}{2}\leq\theta\leq\frac{\pi}{2}$. From this condition we have
\begin{eqnarray}
\frac{2}{\pi}\leq\lambda r\leq1 ,\\
\lambda\ln r+\frac{1}{r}\geq1 .
\end{eqnarray}
Note that under the condition $\frac{\pi}{2}\leq\lambda r\leq1$ there is no solution to $\lambda\ln r+\frac{1}{r}>1$---this two conditions contradict. Thus this statement is proved.

Secondly, we check the assumption---when $\mu^2$ is not real, coupling is not real.

From Eq.10, we can extract the imaginary part of $y_{an}(\mu^2)$, which can be   written as
\begin{multline}
\text{Im}y_{an}(\mu_E^2)=\\
\frac{2}{C_1}\int_0^\infty\frac{\text{Im}(\mu_E^2)}{(x-\text{Re}(\mu_E^2))^2+(\text{Im}(\mu_E^2))^2}\frac{\ud x}{\ln^2|\frac{x}{\Lambda^2}|+\pi^2} .
\end{multline}
It's obvious that $\text{Im}(\mu_E^2)\neq0$ implies $\text{Im}y(\mu_E^2)\neq0$, which leads to the appearance of an imaginary part in $y_{an}(\alpha_{an})$. This can happen only when $\alpha$ is not real number. Thus we finish our check.

\end{document}